\documentclass[11pt]{article}
\usepackage{epsfig}
\usepackage{amsfonts}

\def\laq{~\raise 0.4ex\hbox{$<$}\kern -0.8em\lower 0.62
ex\hbox{$\sim$}~}
\def\gaq{~\raise 0.4ex\hbox{$>$}\kern -0.7em\lower 0.62
ex\hbox{$\sim$}~}

\def\beq{\begin{equation}}
\def\eeq{\end{equation}}
\def\bea{\begin{eqnarray}}
\def\eea{\end{eqnarray}}

\def \ra {\rightarrow}

\def \Da {\Delta}

\def \ga {\gamma}

\def \noi {\noindent}

\begin{document}
\begin{titlepage}

\begin{flushright}
BA-TH/668-13\\
\end{flushright}

\vspace{1.2 cm}

\begin{center}

{\Large{\bf The twin paradox in the presence of gravity}}

\vspace{1cm}

{M. Gasperini}

\bigskip
\normalsize

{\sl Dipartimento di Fisica,
Universit\`a di Bari, \\
Via G. Amendola 173, 70126 Bari, Italy\\
and\\
Istituto Nazionale di Fisica Nucleare, Sezione di Bari, Bari, Italy \\
\vspace{0.3cm}
E-mail: {\tt gasperini@ba.infn.it}
}

\vspace{1cm}

\begin{abstract}
\noi
Conventional wisdom, based on kinematic (flat-space) intuition, tell us that a static twin is aging faster than his traveling twin brother. However, such a situation could be exactly inverted if the two twins are embedded in an external gravitational field, and if the (dynamical)  distortion of the space-time geometry, 
due to gravity, is strong enough to compensate the kinematic effect of the relative twin motion.
\end{abstract}
\end{center}

\bigskip
\begin{center}
---------------------------------------------\\
\vspace {10 mm}
{ Published in\\ {\em Mod. Phys. Lett.} {\bf A 29}, 1450149 (2014)}
\end{center}

\end{titlepage}
 
\newpage
\parskip 0.2cm

It is well known that, after a round trip in Minkowski space, the traveling
twin is {\em younger} than the static twin because of the kinematic effect of time dilatation (see e.g. \cite{1}). It is also well known that gravity affects the ``flow" of proper time with respect to the free flow typical of inertial frames in flat space-time (see e.g. \cite{2}). Not so well known, however, is the fact that if the static twin is at rest in a given gravitational field, while the moving twin travels across regions where the field is weaker, then the time dilatation due to gravity may compensate the kinematic one, and it becomes possible that, after his journey, the traveling twin finds himself {\em older} than his static brother!

In this paper we will illustrate this effect with a simple example in which one of the two twins moves at a constant non-relativistic speed through the weak and static field of a central source. The space-time geometry is described by the metric
\beq
ds^2= (1+2 \phi) dt^2- (1- 2 \phi) |d\vec x|^2,
\label{1}
\eeq
where $\phi=-GM/r$ (we are using units in which $c=1$, and we are working to first order in $\phi$, with $|\phi| \ll1$). The static twin $A$ is at rest at a radial distance $r_1$ from the central body of mass $M$, while the traveling twin $B$ moves away along a radial trajectory from $r_1$ to $r_2>r_1$ and then turns back to $r_1$, at a  speed $v \ll1$ which represents a constant parameter of the motion. We will assume -- as usual in the discussion of the twin paradox -- that the duration of the deceleration/acceleration regimes associated to the ``bounce" occurring at $r_2$ is negligible, i.e. that the sign flip of the radial velocity at $r_2$ may be regarded as instantaneous (so that $|v|=$ const for the whole journey). 

In the absence of gravity ($ \phi \ra 0$) the ratio between the duration of the round trip, referred to the proper-time parameters of the two twins $A$ and $B$, is controlled by the Lorentz factor $\ga$, and is given (as is well known) by:
\beq
{\Da t_A\over \Da t_B}= \ga= {1\over \sqrt{1-v^2}} \simeq 1+{v^2\over 2} >1.
\label{2}
\eeq
In the presence of gravity, however, there is also a gravity-induced distortion of the proper-time duration of the round trip. Such a distortion is due to the combined action of the curved geometry on both time and space intervals (indeed, the proper length of the trip is also modified by gravity). The spatial distortions are the same for both twins, but the time distortions are not, and the overall effect is different for the two twins.

Let us evaluate this effect in the frame of the static twin $A$, at rest at the position $r_1$.  The duration of the trip, referred to the proper-time of twin A (computed within our geometric model, and expanded to first order in $\phi$ and $v^2$), can be expressed as follows:
\bea
\Da \tau_A &=&2 \sqrt{g_{00}(r_1)}\, \Da t_{12}
\nonumber \\ 
&\simeq& 2  (1+\phi_1) \Da t_{12} \simeq {2 \over v} (1+ \phi_1) \int_{r_1}^{r_2} dr \left[ 1 - \phi(r)\right]
\nonumber \\
&\simeq& {2  \over v}\left(r_2-r_1\right)\left(1 -{GM\over r_1}+{GM\over r_2-r_1} \ln {r_2\over r_1} \right), ~~~~~~~~~~~ r_2>r_1.
\label{3}
\eea
Let us then compute, in the same frame, the duration of the trip referred to the proper time of the traveling twin $B$. For the twin B the time-dilatation due to gravity cannot be factorized like in the above equation, since $g_{00}(r)$ varies along the trajectory of the motion. The  duration of the travel for the proper-time of twin $B$ is thus given by:
\bea
\Da \tau_B &\simeq& {2\over v \ga} \int_{r_1}^{r_2} d r\sqrt{g_{00}(r)} \left[ 1 - \phi(r)\right] \simeq {2 \over v \ga}\left(r_2-r_1\right)
\nonumber \\
&\simeq&  {2 \over v}\left(r_2-r_1\right)\left(1- {v^2\over 2}\right).
\label{4}
\eea
Hence:
\beq
{\Da \tau_A\over \Da \tau_B}=1+{v^2\over 2}  -{GM\over r_1}+{GM\over r_2-r_1} \ln {r_2\over r_1} , ~~~~~~~~~~~~ r_2>r_1,
\label{5}
\eeq
to first order in $v^2 \ll1$ and $(GM/r_1) \ll1$, for any value of the end-point of the trip $r_2$, with $r_2>r_1$.

It can be easily checked, now, that the gravitational corrections to the special-relativistic ratio (\ref{2}) satisfy the condition
\beq
-{1\over r_1}+{1\over  r_2-r_1} \ln{r_2\over r_1}  <0, ~~~~~~~~~~  r_2>r_1,
\label{6}
\eeq
and thus contribute with the opposite sign (with respect to the kinematic contribution $v^2/2$) to the mismatch between the two proper times. It follows that even the result $\Da \tau_A <\Da \tau_B$ (namely, a static twin {\em younger} than the traveling one) now becomes possible, provided the parameters of the round trip satisfy the condition
\beq
GM \left( {1\over r_1}- {1\over  r_2-r_1} \ln{r_2\over r_1}\right) >{v^2\over 2}.
\label{7}
\eeq

No such result is possible, instead, if the moving twin travels across regions where the field is {\em stronger} than at the position of his static brother. Let us assume, for instance, that the twin $A$ is at rest at $r=r_2$, while the twin $B$ approaches the central source moving radially from $r_2$ to $r_1<r_2$ and then back to $r_2$, at a constant  speed $v$, as before. By repeating the above computation one finds that Eq. (\ref{5}) is replaced by 
\beq
{\Da \tau_A\over \Da \tau_B}=1+{v^2\over 2}  -{GM\over r_2}+{GM\over r_2-r_1} \ln {r_2\over r_1} , ~~~~~~~~~~~~ r_2>r_1.
\label{8}
\eeq
The gravitational contribution, in this case, satisfies the condition:
\beq
-{1\over r_2}+{1\over  r_2-r_1} \ln{r_2\over r_1}  >0, ~~~~~~~~~~  r_2>r_1.
\label{9}
\eeq
This contribution can only be added (with the same sign) to the kinematic effects, and the net result is always $\Da \tau_A >\Da \tau_B$ (like in the absence of gravity).

Let us now come back to Eq. (\ref{7}), and ask whether such equation may be compatible with the assumed weak-field and non-relativistic approximations. 

\begin{figure}
\centering
\includegraphics[height=6.5 cm]{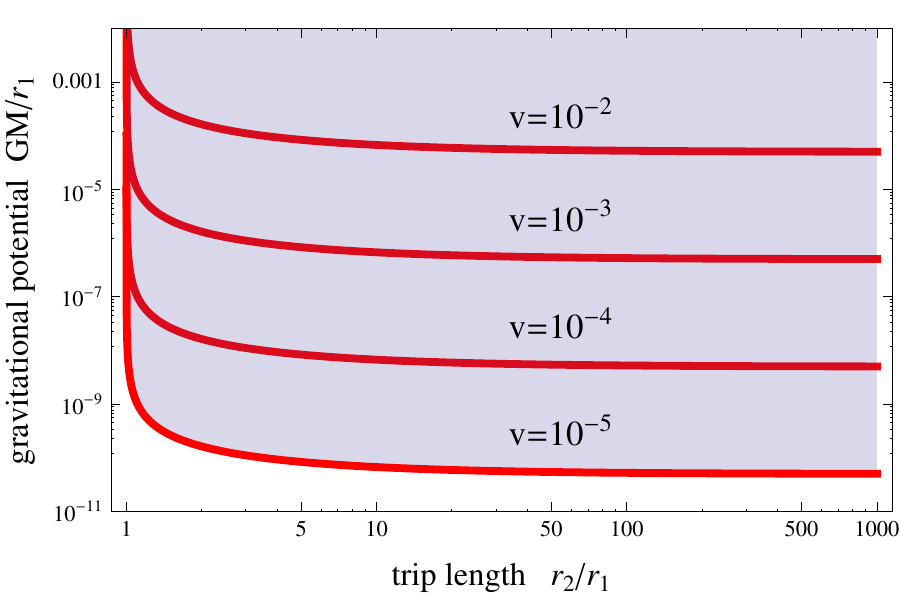}
\caption{Plot of the condition (\ref{7}) for different values of the velocity parameter $v$. For each curve at constant $v$,  the allowed region (the shaded area) lies {\em above} the plotted curve. The figure illustrates the compatibility of Eq. (\ref{7}) with the weak-field ($|\phi_1| \ll 1$) and non-relativistic ($v \ll 1$) approximations.}
\label{Fig1}
\end{figure}

The allowed region determined by the condition (\ref{7}) in the plane $\{x= r_2/r_1$, $y= GM/r_1 \}$ is illustrated in Fig. \ref{Fig1}, for various values of the parameter $v$ ranging from $10^{-5}$ to $10^{-2}$. For each curve $v=$ const, the allowed region (the shaded area) lies {\em above} the plotted curve. As evident from the figure, higher values of the gravitational potential $|\phi_1|=GM/r_1$ are needed to compensate the effects of higher velocities. However, for any given non-relativistic value of $v$, we can always find a a gravitational field which is weak enough to be described in the linear approximation ($|\phi_1| \ll 1$), and strong enough to keep the static twin younger than than his traveling brother, $\Da \tau_A <\Da \tau_B$ (provided the end-point of the trip is sufficiently far away from the initial position, $r_2 \gg r_1$). 

The above results can be easily generalized to the case of strong gravitational fields and relativistic velocities by considering, for instance, two twins embedded in the Schwarzschild geometry, described by the metric:
\beq
ds^2= \left(1-{2m\over r} \right) dt^2- \left(1-{2m\over r} \right)^{-1} dr^2 - r^2 \left( d\theta^2+ \sin^2 \theta d\varphi^2\right).
\label{10}
\eeq
Comparing, as before, the proper time of the twin $A$ (at rest at $r=r_1$) with that of the twin $B$ (traveling from $r_1$ to $r_2>r_1$ and then back to $r_1$) we obtain:
\beq
{\Da \tau_A\over \Da \tau_B}=  \left(1-{2m\over r_1} \right)^{1/2}{\ga\over r_2-r_1 } \int_{r_1}^{r_2} d r  \left(1-{2m\over r} \right)^{-1/2}, ~~~~~~~~  r_2>r_1.
\label{11}
\eeq
Computing the above integral we then find that the static twin $A$ is aging {\em less} than his traveling brother $B$ (i.e., that $\Da \tau_A <\Da \tau_B$ ), for $r_2>r_1$, provided that:
\bea 
&& 
{\sqrt{1-{2m\over r_1}}\over r_2-r_1}\left[r_2 \sqrt{1-{2m\over r_2}} -
r_1 \sqrt{1-{2m\over r_1}} + m \ln \left( r_2\sqrt{1-{2m\over r_2}} +r_2-m \over 
r_1\sqrt{1-{2m\over r_1}} +r_1-m \right) \right] <
\nonumber \\ &&
~~~~~~~~~~~~~<\sqrt{1-v^2}.
\label{12}
\eea
In the non-relativistic limit, and for weak-enough fields, we may easily recover from this equation the condition previously reported in Eq. (\ref{7}).

\begin{figure}
\centering
\vspace{0.5 cm}
\includegraphics[height=6.5 cm]{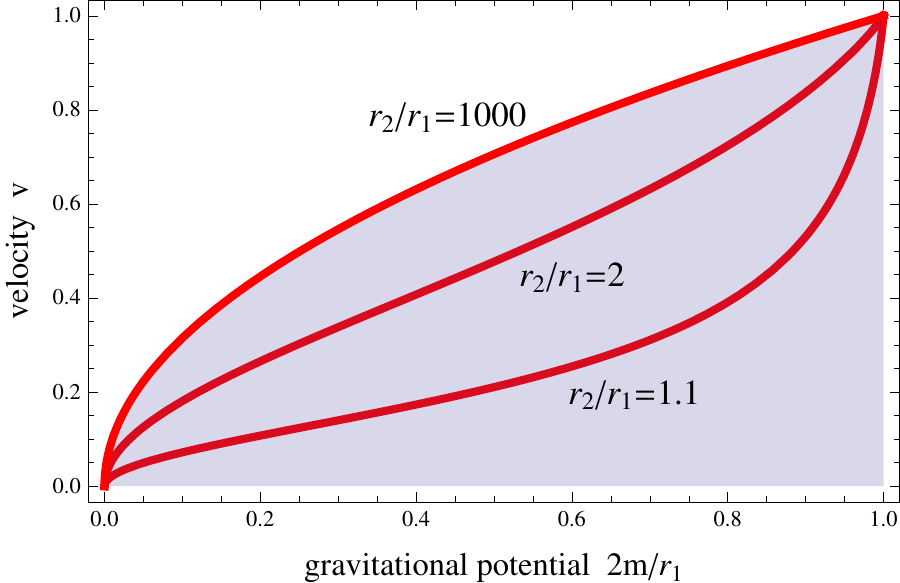}
\caption{Plot of the condition (\ref{12}) for different values of the parameter $r_2/r_1$. For each curve at $r_2/r_1=$ const, the allowed region (the shaded area) lies {\em below} the plotted curve. In the limit $v \ra 1$ the condition (\ref{12}) can  be satisfied only for $r_1 \ra 2m$.}
\label{Fig2}
\end{figure}

The round trip configurations satisfying the above inequality define, for any given value of $r_2/r_1$, an allowed region in the two-dimensional parameter space spanned by the coordinates $x=2m/r_1$ and $y=v$ (both ranging from $0$ to $1$). Such an allowed region is illustrated in Fig. \ref{Fig2}, for each curve plotted at constant values of $r_2/r_1$, as the shaded portion of the $\{x,y\}$ plane lying {\em below} the given curve. The figure -- mainly concentrated on the illustration of the strong-field/relativistic regime -- clearly shows that, as the velocity of the traveling twin $B$ tends to $c$, the radial position $r_1$ of the static twin $A$ must approach the Schwarzschild horizon $r_1=2m$, if he wants to avoid aging more than his traveling brother. 

We can check, finally, that for $r_2 \gg r_1$ the allowed region rapidly saturates the portion of  $\{x,y\}$  plane bounded from above by the curve $y <\sqrt{x}$ (i.e. $v <\sqrt{2m/r_1}$), obtained from Eq. (\ref{12}) in the limit $r_2/r_1 \ra \infty$. In Fig. \ref{Fig2} such an upper bound practically coincides with the curve plotted for $r_2/r_1= 10^3$. We can also check, however, that for any given value of $v$ and of $r_2/r_1$ it is always possible to find a position $r_1$ such that $\Da \tau_A <\Da \tau_B$. This extends to strong fields  the previous results obtained in the weak-field approximation, thus confirming the unique ``anti-aging" virtue of the gravitational field. 

Let us hope that future technology will be able to apply this exceptional virtue of gravity to keep us young, as long as possible!




\begin{thebibliography}{99.}
\newcommand{\bb}{\bibitem}

\bibitem{1}
Resnick, R. (1968) Introduction to Special Relativity. John Wiley \& Sons, New York.

\bb{2}Weinberg, S. (1972) Gravitation and Cosmology. John Wiley \& Sons, New York.


\end{thebibliography}
\end{document}